\documentclass{article}
\usepackage{printlen}
\usepackage{authblk}
\usepackage{graphicx} 
\usepackage{color, colortbl}
\usepackage{geometry}
\usepackage{lmodern}
\usepackage[size=tiny]{todonotes}
\usepackage{multirow}
\usepackage{float}
\usepackage{adjustbox}
\usepackage{subcaption}
\usepackage{hyperref}
\usepackage{comment}
\usepackage{array}
\usepackage{amsmath,amssymb}
\usepackage{csvsimple} 
\usepackage{hyperref}
\usepackage{longtable} 
\usepackage{arydshln}
\usepackage{placeins}
\usepackage[numbers]{natbib}
\usepackage{tabularx, booktabs}
\usepackage{makecell}
\usetikzlibrary{calc, arrows.meta, intersections, patterns, positioning, shapes.misc, fadings, through,decorations.pathreplacing}

\definecolor{ColorOne}{named}{blue}
\definecolor{ColorTwo}{named}{orange}
\definecolor{ColorThree}{named}{purple}



\newcolumntype{P}[1]{>{\centering\arraybackslash}p{#1}}
\newcolumntype{M}[1]{>{\centering\arraybackslash}m{#1}}
\newcommand{\arxivlargepp}[0]{\texttt{arxiv-09/23}}
\newcommand{\arxivsmallpp}[0]{\texttt{arxiv-09/23-top40}}
\newcommand{\arxivlargeppp}[0]{\texttt{arxiv-06/23}}
\newcommand{\arxivsmallppp}[0]{\texttt{arxiv-06/23-top40}}
\newcommand{\arxivlargep}[0]{\texttt{arxiv-01/24}}
\newcommand{\arxivsmallp}[0]{\texttt{arxiv-01/24-top40}}
\newcommand{\arxivlarge}[0]{\texttt{arxiv-09/24}}
\newcommand{\arxivsmall}[0]{\texttt{arxiv-09/24-top40}}

\newcommand{\se}[1]{\textcolor{black}{#1}}
\newcommand{\cl}[1]{\textcolor{black}{#1}}

\title{
NLLG Quarterly arXiv Report 09/24: 
\\
What are the most influential current 
AI 
Papers?
}
\author[1]{Christoph Leiter} 
\author[1]{Jonas Belouadi}
\author[1,3]{Yanran Chen} 
\author[1]{Ran Zhang}
\author[1]{Daniil Larionov}
\author[2,1]{Aida Kostikova}
\author[1,3]{Steffen Eger} 
\affil[1]{Natural Language Learning \& Generation (NLLG)\footnote{\url{https://nl2g.github.io/}}, University of Mannheim}
\affil[2]{Knowledge Representation and Machine Learning, 
Bielefeld University}
\affil[3]{University of Technology Nuremberg}
\date{}

\begin{document}

\maketitle


\begin{abstract}
  The NLLG (Natural Language Learning \& Generation) arXiv reports assist in
  navigating the rapidly evolving landscape of NLP and AI research across
  cs.CL, cs.CV, cs.AI, and cs.LG categories. This fourth installment captures a
  transformative period in AI history---from January 1, 2023, following
  ChatGPT's debut, through September 30, 2024. Our analysis reveals substantial
  new developments in the field---with 45\% of the top 40 most-cited papers
  being new entries since our last report \se{eight months ago}---and offers 
  insights
  into emerging trends and major breakthroughs, \se{such as novel multimodal architectures, including diffusion and state space models}. \se{Natural Language Processing (NLP; cs.CL) remains the dominant main category in the list of our top-40 papers but its dominance is on the decline in favor of Computer vision (cs.CV) and general machine learning (cs.LG).}
  This report also presents novel findings on the integration of generative AI
  in academic writing, documenting its increasing adoption since 2022 while
  revealing an intriguing pattern: top-cited papers show notably fewer markers
  of AI-generated content compared to random samples. Furthermore, we track the
  evolution of AI-associated language, identifying declining trends in
  previously common indicators such as ``delve''. 
\end{abstract}
\section{Introduction}
Keeping up with the rapid advancements in Artificial Intelligence (AI) is
becoming increasingly challenging for both researchers and professionals. 
\se{Today}, 
more scientific preprints are published than ever before, accelerating the
pace of progress. In this environment, researchers and professionals who wish
to stay informed about the latest developments often 
\se{cannot} 
wait for
publications in journals and conference proceedings, as the information can
quickly become outdated.

The NLLG Quarterly \se{arXiv} reports \cite{eger2023nllg, zhang2023nllg, chen2024nllg}
produced by the Natural Language Learning \& Generation (NLLG;\@
\url{https://nl2g.github.io/}) \se{Lab} take a different approach by exploring and
presenting cutting-edge papers based on their immediate impact in the
scientific community---specifically, through the analysis of time-normalized
citation counts. This \se{current} report is the fourth in the series and continues to
offer a comprehensive examination of the most influential AI papers published
on arXiv, 
this time between January 1, 2023, and
September 30, 2024. 

By analyzing the top-cited papers, our goal is to highlight
key trends and breakthroughs  
shaping the future of AI.\@ We focus on
arXiv due to its critical role as a preprint server for disseminating
cutting-edge research, particularly within the computer science
community~\cite{Eger2018PredictingRT,Clement2019OnTU}.

\se{Besides presenting the list of top-cited papers in the given time frame, each report has a particular focus of analysis, contrasting top papers with random arXiv papers.  
While our previous reports focused on (1) the geographic and institution-related affiliation of top papers, finding that Europe and academia are underrepresented compared to random papers, and (2) the writing quality of top papers, finding that they are better written according to various criteria than random papers, the current report focuses on AI generated content in top papers vs.\ random papers. This is an important research question as LLM generated content may correlate with writing quality which in turn might affect citation counts, perhaps explaining in part the high ranks of some papers.}

The key insights from this report are: (i) 
\se{AI} 
research continues to focus on
large language models \se{(LLMs)} but is increasingly exploring alternatives to the
transformer architecture, including diffusion and state space
models \cite{zhou2024transfusionpredicttokendiffuse,zhu2024visionmambaefficientvisual}; \se{ 
(ii) while the field of Natural Language Processing (NLP; cs.CL) continues to dominate the top papers, its importance, in terms of papers within the top-40, has successively decreased over the course of our previous reports. Computer vision (cs.CV) and more general machine learning (cs.LG) are more strongly represented, in turn; 
this may indicate a shift towards broader and more diversified (multimodal) architectures.}
(iii) While the use of LLMs to assist in writing is on the rise overall, our
analysis of the top-40 papers reveals that their use remains surprisingly low
in this subset, suggesting a potential inverse correlation between the use of
LLMs and the overall quality of research conducted. \se{We also show that certain keywords that were previously indicative of LLM usage, such as \emph{delve}, are already decreasing in frequency again, perhaps suggesting that newer LLMs have other vocabulary preferences.}
We make the code and data artifacts produced for the analysis in this work publicly available at
\url{https://github.com/NL2G/Quarterly-Arxiv}. %

\FloatBarrier%
\section{Methods}\label{sec:methodology}
As in our previous arXiv reports \cite{eger2023nllg,zhang2023nllg,chen2024nllg}, we use week-based normalized z-scores of citation counts to identify to identify the most influential papers for the categories AI, CV, CL and LG. Here, we briefly summarize the setup:

\begin{enumerate}
\item \textbf{Data Retrieval}: 
    We use an arXiv Python API\footnote{\url{https://github.com/lukasschwab/arxiv.py}} to download metadata of all papers belonging to any of the arXiv categories cs.CL, cs.AI, cs.CV and cs.LG that were published from 2023/01/01 to 2024/09/30. The citation counts are dependent on the retrieval date: \textbf{2024/11/20}. We refer to this new dataset as \arxivlarge{}.
    
\item \textbf{Paper Ranking}: 
    As in our previous reports, we use a SemanticScholar Python API\footnote{\url{https://github.com/danielnsilva/semanticscholar}} to determine the citation counts of each paper in \arxivlarge{}. Next, for each paper's citation count, we calculate the normalized z-score compared to all papers that were published in the same week:\footnote{We average these z-scores across each possible week start; Monday, Tuesday, etc..}

\begin{align}
    z_{\se{i}(t)} &= \frac{c_{\se{i}(t)}-\textit{mean}(\mathbf{c}(t))}{\textit{std}(\mathbf{c}(t))} \label{eq:eq1} \\
    \hat{z}_{i\se{(t)}} &= \textit{mean}(\mathbf{z}_{i(t^d)}) - \textit{std}(\mathbf{z}_{i(t^d)}) \label{eq:eq2}
 \end{align}
 for paper $i$ published in week $t$ with citation count $c_{\se{i}(t)}$; $\mathbf{c}(t)$ is the list of citation counts of all papers published in week $t$. For more details see \cite{zhang2023nllg}. Sorting by these z-scores allows us to fairly select the top 40 papers throughout a time period.

\begin{figure}[tb]
    \centering
    \begin{subfigure}[t]{0.49\textwidth}
    \includegraphics[width=\linewidth]{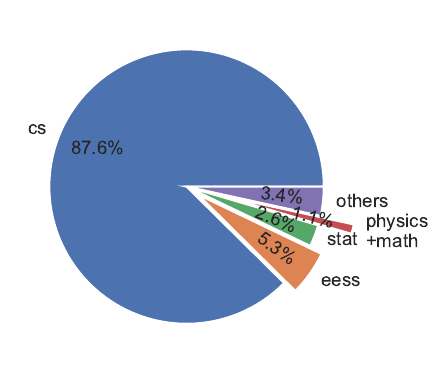}
    \caption{Top-level category distribution. The categories math and physics are merged together.}
    \label{fig:top_cat_dis}
    \end{subfigure}\hspace{.1cm}
    \begin{subfigure}[t]{0.495\textwidth}
    \includegraphics[width=\linewidth]{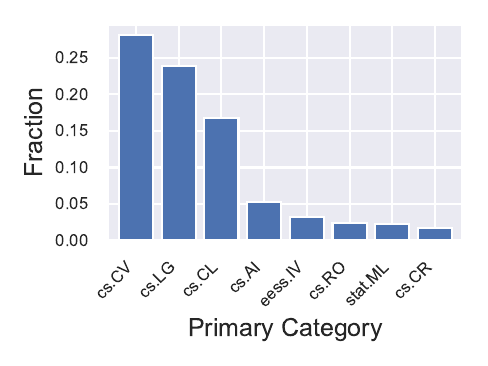}
    \caption{Top 8 sub-category distribution in our dataset.}
    \label{fig:fine_cat}
    \end{subfigure}
    \caption{Distribution of top-level and sub-categories}
    \label{fig:enter-label}
\end{figure}

\item \textbf{Manual Evaluation\se{:}}
We manually confirm the top-40 papers and remove those that were published before their arXiv release. As these papers already gathered citations outside of arXiv, leaving them in would introduce bias. We determine these prior releases based on web searches. We refer to the dataset of top-40 papers as \arxivsmall{}.
\end{enumerate}
\FloatBarrier%
\section{Data Exploration}\label{dataset_stat}
In this section, we explore our dataset to highlight the overall trends in numbers and citation counts since 2023/1/1, notably  
\se{discussing} 
the months from 2024/01/31 
\se{(our last report)} 
to 2024/09/30.

\begin{table}[tb]
    \centering
    \small
    \begin{tabular}{ccccc}
         \toprule
         \textbf{Dataset name} & \textbf{Size} & \textbf{Time period} & \textbf{Categories} & \textbf{Search Query}\\ 
         \midrule
         \arxivlarge & 127,127 & \multirow{2}{*}{2023/01/01--2024/09/30} & 139 & \multirow{2}{*}{cs.CL, cs.LG, cs.AI, cs.CV}\\ 
         \arxivsmall &  40 &  & 6\\ \midrule 
         \arxivlargep &  71,139 & \multirow{2}{*}{2023/01/01--2024/01/31} & 135 & \multirow{2}{*}{cs.CL, cs.LG, cs.AI, cs.CV}\\ 
         \arxivsmallp &  40 &  & 6\\ \midrule
         \arxivlargepp &  47\se{,}331 & \multirow{2}{*}{2023/01/01--2023/09/30} & 130 & \multirow{2}{*}{cs.CL, cs.LG, cs.AI, cs.CV}\\
         \arxivsmallpp &  40 & & 6 &\\ \midrule
         \arxivlargeppp &  20,843 & \multirow{2}{*}{2023/01/01--2023/06/30} & 123 & \multirow{2}{*}{cs.CL, cs.LG}\\
         \arxivsmallppp &  40 & & 5\\ \bottomrule
    \end{tabular}
    \caption{This table gives an overview of the datasets that we create in this and our previous reports. The colum \textit{Size} indicates the number of data samples. The column \textit{time period} shows the time frame covered in the dataset. \textit{Categories} shows the number of unique primary categories and \textit{search query} shows, which arXiv categories where included in the API search request.}
    \label{table:stats}
\end{table}

\paragraph{Data Distribution} Table \ref{table:stats} shows how this report's data sets compare to the previous reports. Compared to the previous report, the number of papers almost doubled. This is expected as this report covers a larger time span than each of the previous ones.  
ArXiv categories are composed of a top-level category, e.g.\ cs for computer science and a sub-category, e.g.\ CL for  
natural language processing 
\se{(computation \& language)}.\footnote{See \url{https://arxiv.org/category_taxonomy}.} Figure \ref{fig:top_cat_dis} shows the distribution of top-level categories in our data set. Compared to our 
\se{last} 
report, there are no big changes. The most notable differences are \textit{physics+math} dropping from $2.2\%$ to $1.1\%$, \textit{cs} gaining $0.6\%$ and \textit{others} gaining $0.7\%$. Further, Figure \ref{fig:fine_cat} shows the distribution of the 8 most common sub-categories. Here, the most notable difference to the previous report is \textit{cs.RO} (Robotics) overtaking \textit{stat.ML}. \se{Overall, Computer Vision (cs.CV) and Machine Learning (cs.LG) dominate with above 20\% of all papers, Natural Language Processing (cs.CL) is third.}

\begin{figure}[tb]
    \centering
    \begin{subfigure}[t]{0.49\textwidth}
    \includegraphics[width=\linewidth]{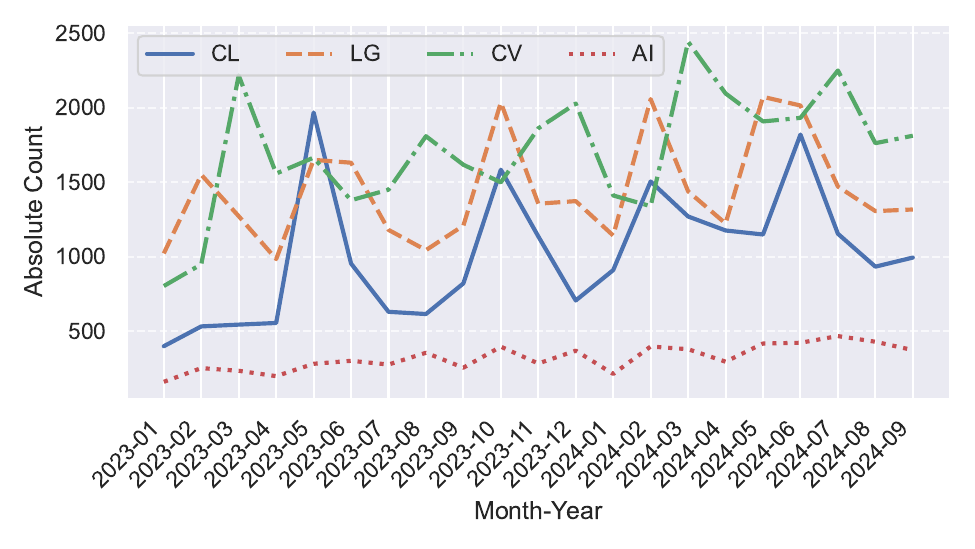}
    \caption{Absolute publication counts per primary category per month..}
    \label{fig:abs_dist}
    \end{subfigure}\hspace{.1cm}
    \begin{subfigure}[t]{0.49\textwidth}
    \includegraphics[width=\linewidth]{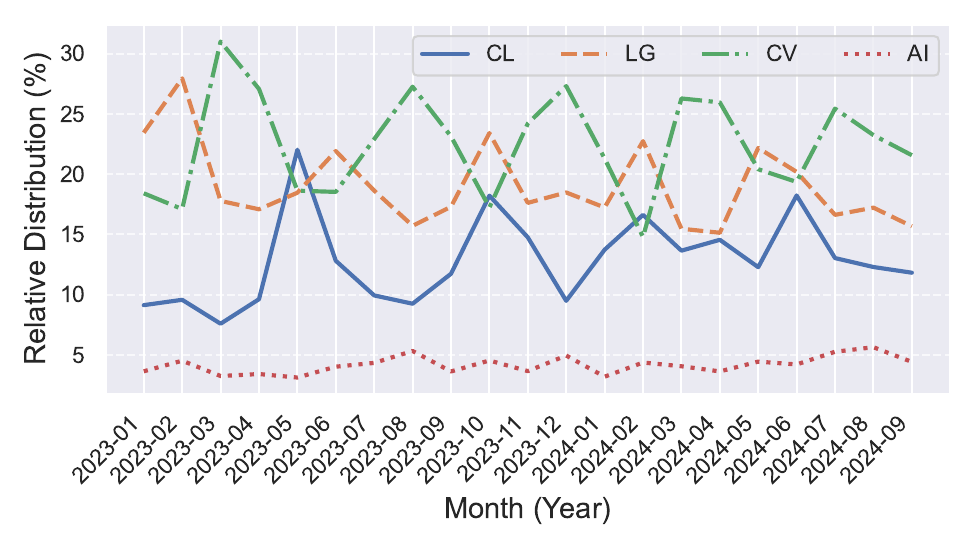}
    \caption{Relative publication counts per primary category per month.}
    \label{fig:rel_dist}
    \end{subfigure}
    \caption{Absolute and relative citation counts.}
    \label{fig:current_prev_top}
\end{figure}

  \paragraph{Publication counts over time} %
  Next, we analyze the publication counts per month per search \se{Arxiv} category.\footnote{As compared to the previous reports with week-wise plots, we choose a month-wise plot to be less prone to outliers and because this new report covers a larger time span.} Figure \ref{fig:abs_dist} shows the absolute number of publications per month per primary category. 
  We can see that the numbers are slowly increasing over the months. Like in the previous reports, the most 
  \se{frequent} 
  category is cs.CV, followed by cs.LG and cs.CL. Also, each category has a wave-like pattern with high-publication months followed by low-publication months. Our previous reports speculate that this is caused by conference schedules. To further visualize these patterns, Figure \ref{fig:rel_dist} shows the relative publication counts per month, i.e., each absolute number is divided by the month's total. Interestingly, it seems that cs.CV is inverted to cs.CL and cs.LG. That means, cs.CV publications are higher when cs.CL and cs.LG publications are lower and vice versa; \se{and} all of 
  \se{them}
  follow 
  a 
  \se{cyclic}
  pattern. A third interpretation would be that the research progress is slowing down,  
but given the high number of new top40 papers (see Section~\ref{sec:top_papers}) this seems unlikely. Future work may further investigate this. 

\begin{figure}[tb]
    \centering
    \includegraphics[width=0.95\linewidth]{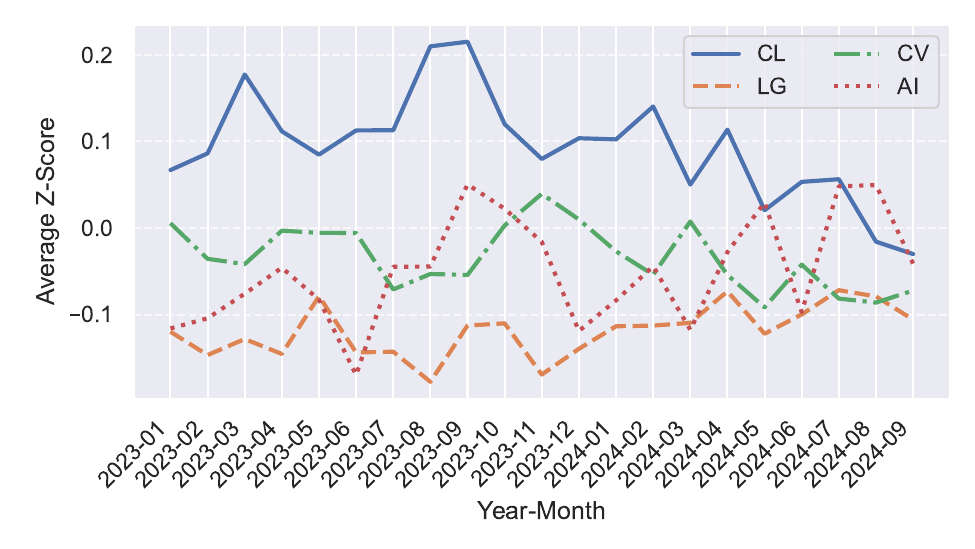}
    \caption{Average z-score per month.}
    \label{fig:z-score}
\end{figure}

\paragraph{Z-Scores over time}
With respect to citation counts, we analyze the average z-score per month (see Figure \ref{fig:z-score}). For each month the cs.CL papers are the most influential, except August 2024, where cs.AI has the highest scores. Also, the plots seem to converge 
over the recent months, \cl{towards a z-score of -0.05}. This might unveil a small bias that is introduced by highly influential older papers receiving more citations faster than less cited papers published in the same week or highly influential newer papers. Another cause might be the higher volume of papers in the more recent months.

\FloatBarrier%
\section{Top-40 papers}
\label{sec:top_papers}

Tables \ref{tab:top20} and \ref{tab:top2140} show the top 20 and top 21-40 papers of the time period from 2023/1/1 to 2024/9/30. Since our last report, 11 newly published papers were included into our list, including all of the top 3 papers.\footnote{This time, we also noticed a bug in SemanticScholar attributing over 900 citations to the paper \cite{Biswas2024IntelligentCD}, which would make it the most influential paper on our list. However, most citations are actually listed as references in the citing paper, but merely published in the same journal. For this reason, we exclude this paper from our list.} Additionally, 7 papers that were released before 2024/1/30 but not listed in our previous report are now included. Therefore, 45\% of the top \cl{40} 
papers have changed compared to our last report. Most of the newly included papers introduce new foundation models, e.g.\ Qwen2 \cite{yang2024qwen2technicalreport}, a family of multi-lingual open-source LLMs by Alibaba; LLaMA 3 \cite{grattafiori2024llama3herdmodels} a family of open source LLMs by Meta; Phi3 \cite{abdin2024phi3technicalreporthighly}, a family of small open-source LLMs by Microsoft; Gemini 1.5 \cite{geminiteam2024gemini15unlockingmultimodal} a family of closed-source LLMs by Google; DeepSeek-V2 \cite{deepseekai2024deepseekv2strongeconomicalefficient}, a Mixture-of-Experts LLM by Deepseek-AI; and ChatGLM \cite{glm2024chatglmfamilylargelanguage}, a family of Chinese and English trained open-source LLMs by Zhipu-AI.
Additionally, some works consider new architectural and training paradigms. For example, Transfusion \cite{zhou2024transfusionpredicttokendiffuse} combines diffusion with next-token prediction to train multi-modal transformer models. Kolmogorov-Arnold Networks \cite{liu2024kankolmogorovarnoldnetworks} offer an alternative to multi-layer perceptrons; \se{an}d Vision Mamba \cite{zhu2024visionmambaefficientvisual} leverages the recently proposed Mamba architecture \cite{gu2023mamba}
in a computer vision setting. 

\definecolor{Gray}{gray}{0.7}
\definecolor{LGray}{gray}{0.9}

\begin{table}[tbp]
    \centering
    \scriptsize
    \setlength{\tabcolsep}{1pt}
    \renewcommand{\arraystretch}{0.1}
    \begin{tabular}{m{0.5cm} m{5.4cm} M{0.5cm} M{2.83cm} M{1.4cm} M{0.6cm} M{1cm} M{0.7cm} M{0.7cm} M{0.7cm}}
    \toprule
\textbf{No.} & \textbf{Title} & \textbf{Cat.} & \textbf{Link} & \textbf{Week} & \textbf{Cit.} & \textbf{Z-score} & \textbf{vs.\ 1/24} & \textbf{vs.\ 9/23} & \textbf{vs.\ 6/23} \\ \midrule
\rowcolor{lime} 1 & Qwen2 Technical Report & CL & \url{http://arxiv.org/abs/2407.10671} & 24/07/14-24/07/20 & 320 & 38.4 & - & - & - \\ \midrule
\rowcolor{lime} 2 & The Llama 3 Herd of Models & AI & \url{http://arxiv.org/abs/2407.21783} & 24/07/28-24/08/03 & 1192 & 35.4 & - & - & - \\ \midrule
3 & GPT-4 Technical Report & CL & \url{http://arxiv.org/abs/2303.08774} & 23/03/12-23/03/18 & 8670 & 35.3 & $\downarrow$2 & $\downarrow$2 & $\downarrow$1 \\ \midrule
4 & Judging LLM-as-a-Judge with MT-Bench and Chatbot Arena & CL & \url{http://arxiv.org/abs/2306.05685} & 23/06/04-23/06/10 & 2551 & 34.5 & $\rightarrow$ & $\uparrow$1 & $\uparrow$7 \\ \midrule
\rowcolor{LGray} 5 & Llama 2: Open Foundation and Fine-Tuned Chat Models & CL & \url{http://arxiv.org/abs/2307.09288} & 23/07/16-23/07/22 & 8402 & 34.3 & $\downarrow$3 & $\downarrow$3 & - \\ \midrule
\rowcolor{Gray} 6 & Mamba: Linear-Time Sequence Modeling with Selective State Spaces & LG & \url{http://arxiv.org/abs/2312.00752} & 23/11/26-23/12/02 & 1280 & 33.6 & $\uparrow$3 & - & - \\ \midrule
7 & Direct Preference Optimization: Your Language Model is Secretly a Reward Model & LG & \url{http://arxiv.org/abs/2305.18290} & 23/05/28-23/06/03 & 2025 & 33.6 & $\uparrow$7 & $\uparrow$15 & - \\ \midrule
8 & LLaMA: Open and Efficient Foundation Language Models & CL & \url{http://arxiv.org/abs/2302.13971} & 23/02/26-23/03/04 & 9170 & 33.5 & $\downarrow$6 & $\downarrow$6 & - \\ \midrule
\rowcolor{lime} 9 & Phi-3 Technical Report: A Highly Capable Language Model Locally on Your Phone & CL & \url{http://arxiv.org/abs/2404.14219} & 24/04/21-24/04/27 & 479 & 31.7 & - & - & - \\ \midrule
\rowcolor{Gray} 10 & Retrieval-Augmented Generation for Large Language Models: A Survey & CL & \url{http://arxiv.org/abs/2312.10997} & 23/12/17-23/12/23 & 777 & 31.3 & $\uparrow$N & - & - \\ \midrule
\rowcolor{lime} 11 & YOLOv9: Learning What You Want to Learn Using Programmable Gradient Information & CV & \url{http://arxiv.org/abs/2402.13616} & 24/02/18-24/02/24 & 404 & 30.3 & - & - & - \\ \midrule
12 & BLIP-2: Bootstrapping Language-Image Pre-training with Frozen Image Encoders and Large Language Models & CV & \url{http://arxiv.org/abs/2301.12597} & 23/01/29-23/02/04 & 2977 & 29.6 & $\downarrow$5 & $\downarrow$6 & $\downarrow$6 \\ \midrule
13 & Segment Anything & CV & \url{http://arxiv.org/abs/2304.02643} & 23/04/02-23/04/08 & 4496 & 29.6 & $\downarrow$5 & $\downarrow$6 & $\downarrow$3 \\ \midrule
\rowcolor{LGray} 14 & 3D Gaussian Splatting for Real-Time Radiance Field Rendering & GR & \url{http://arxiv.org/abs/2308.04079} & 23/08/06-23/08/12 & 1811 & 29.5 & $\uparrow$19 & - & - \\ \midrule
15 & Sparks of Artificial General Intelligence: Early experiments with GPT-4 & CL & \url{http://arxiv.org/abs/2303.12712} & 23/03/19-23/03/25 & 2494 & 29.3 & $\downarrow$10 & $\downarrow$11 & $\downarrow$12 \\ \midrule
\rowcolor{lime} 16 & Gemini 1.5: Unlocking multimodal understanding across millions of tokens of context & CL & \url{http://arxiv.org/abs/2403.05530} & 24/03/03-24/03/09 & 682 & 29.0 & - & - & - \\ \midrule
\rowcolor{Gray} 17 & Improved Baselines with Visual Instruction Tuning & CV & \url{http://arxiv.org/abs/2310.03744} & 23/10/01-23/10/07 & 1503 & 28.0 & $\uparrow$N & - & - \\ \midrule
18 & QLoRA: Efficient Finetuning of Quantized LLMs & LG & \url{http://arxiv.org/abs/2305.14314} & 23/05/21-23/05/27 & 1648 & 27.6 & $\downarrow$8 & $\downarrow$7 & $\downarrow$9 \\ \midrule
\rowcolor{LGray} 19 & Code Llama: Open Foundation Models for Code & CL & \url{http://arxiv.org/abs/2308.12950} & 23/08/20-23/08/26 & 1349 & 27.3 & $\downarrow$8 & - & - \\ \midrule
\rowcolor{lime} 20 & Qwen2-VL: Enhancing Vision-Language Model's Perception of the World at Any Resolution & CV & \url{http://arxiv.org/abs/2409.12191} & 24/09/15-24/09/21 & 88 & 27.0 & - & - & - \\ 
\\ \bottomrule
\end{tabular}
    \caption{\textbf{Top 20} papers with their titles, primary categories (cs.XX), links, publication weeks, citation counts, z-scores, and trends in rankings compared to the top-40 lists in our previous reports \protect\cite{zhang2023nllg}, 
    ranked by z-scores as of January 21, 2024. Row colors correspond to the publication time: \colorbox{white}{white} (2023/1/1-2023/6/30, added in \protect\arxivlargeppp{}), \colorbox{LGray}{light gray} (2023/7/1-2023/9/30, added in \protect\arxivlargepp{}), \colorbox{Gray}{gray} (2023/10/1-2024/1/31, added in \protect\arxivlargep{}), and \colorbox{lime}{lime} (2024/1/31-2024/9/30, added in \protect\arxivlarge{}, i.e., this report).}
    \label{tab:top20}
\end{table}

\definecolor{Gray}{gray}{0.7}
\definecolor{LGray}{gray}{0.9}

\begin{table}[tbp]
    \centering
    \scriptsize
    \setlength{\tabcolsep}{1pt}
    \renewcommand{\arraystretch}{0.1}
    \begin{tabular}{m{0.5cm} m{5.4cm} M{0.5cm} M{2.83cm} M{1.4cm} M{0.6cm} M{1cm} M{0.7cm} M{0.7cm} M{0.7cm}}
    \toprule
\textbf{No.} & \textbf{Title} & \textbf{Cat.} & \textbf{Link} & \textbf{Week} & \textbf{Cit.} & \textbf{Z-score} & \textbf{vs.\ 1/24} & \textbf{vs.\ 9/23} & \textbf{vs.\ 6/23} \\ \midrule
\rowcolor{Gray} 21 & Mistral 7B & CL & \url{http://arxiv.org/abs/2310.06825} & 23/10/08-23/10/14 & 1323 & 26.4 & $\uparrow$3 & - & - \\ \midrule
\rowcolor{Gray} 22 & Mixtral of Experts & LG & \url{http://arxiv.org/abs/2401.04088} & 24/01/07-24/01/13 & 648 & 25.4 & $\uparrow$1 & - & - \\ \midrule
23 & Adding Conditional Control to Text-to-Image Diffusion Models & CV & \url{http://arxiv.org/abs/2302.05543} & 23/02/05-23/02/11 & 2738 & 25.4 & $\downarrow$11 & $\downarrow$8 & - \\ \midrule
\rowcolor{lime} 24 & ChatGLM: A Family of Large Language Models from GLM-130B to GLM-4 All Tools & CL & \url{http://arxiv.org/abs/2406.12793} & 24/06/16-24/06/22 & 150 & 25.3 & - & - & - \\ \midrule
25 & Efficient Memory Management for Large Language Model Serving with PagedAttention & LG & \url{http://arxiv.org/abs/2309.06180} & 23/09/10-23/09/16 & 971 & 24.4 & $\rightarrow$ & - & - \\ \midrule
26 & Visual Instruction Tuning & CV & \url{http://arxiv.org/abs/2304.08485} & 23/04/16-23/04/22 & 2611 & 24.2 & $\downarrow$11 & $\downarrow$14 & $\downarrow$19 \\ \midrule
27 & InstructBLIP: Towards General-purpose Vision-Language Models with Instruction Tuning & CV & \url{http://arxiv.org/abs/2305.06500} & 23/05/07-23/05/13 & 1409 & 23.8 & $\downarrow$8 & $\downarrow$6 & $\uparrow$6 \\ \midrule
\rowcolor{LGray} 28 & Qwen Technical Report & CL & \url{http://arxiv.org/abs/2309.16609} & 23/09/24-23/09/30 & 960 & 23.1 & $\uparrow$N & - & - \\ \midrule
\rowcolor{Gray} 29 & DeepSeek-Coder: When the Large Language Model Meets Programming -- The Rise of Code Intelligence & SE & \url{http://arxiv.org/abs/2401.14196} & 24/01/21-24/01/27 & 356 & 21.6 & $\uparrow$N & - & - \\ \midrule
\rowcolor{lime} 30 & DeepSeek-V2: A Strong, Economical, and Efficient Mixture-of-Experts Language Model & CL & \url{http://arxiv.org/abs/2405.04434} & 24/05/05-24/05/11 & 152 & 21.3 & - & - & - \\ \midrule
\rowcolor{LGray} 31 & Baichuan 2: Open Large-scale Language Models & CL & \url{http://arxiv.org/abs/2309.10305} & 23/09/17-23/09/23 & 552 & 21.1 & $\downarrow$14 & $\uparrow$6 & - \\ \midrule
\rowcolor{LGray} 32 & Universal and Transferable Adversarial Attacks on Aligned Language Models & CL & \url{http://arxiv.org/abs/2307.15043} & 23/07/23-23/07/29 & 881 & 20.0 & $\uparrow$N & - & - \\ \midrule
33 & PaLM-E: An Embodied Multimodal Language Model & LG & \url{http://arxiv.org/abs/2303.03378} & 23/03/05-23/03/11 & 1214 & 19.9 & $\downarrow$17 & $\downarrow$23 & $\downarrow$28 \\ \midrule
\rowcolor{LGray} 34 & SDXL: Improving Latent Diffusion Models for High-Resolution Image Synthesis & CV & \url{http://arxiv.org/abs/2307.01952} & 23/07/02-23/07/08 & 1224 & 19.2 & $\downarrow$4 & - & - \\ \midrule
35 & Tree of Thoughts: Deliberate Problem Solving with Large Language Models & CL & \url{http://arxiv.org/abs/2305.10601} & 23/05/14-23/05/20 & 1140 & 18.8 & $\downarrow$13 & $\downarrow$21 & $\downarrow$22 \\ \midrule
\rowcolor{lime} 36 & Length-Controlled AlpacaEval: A Simple Way to Debias Automatic Evaluators & LG & \url{http://arxiv.org/abs/2404.04475} & 24/03/31-24/04/06 & 175 & 18.3 & - & - & - \\ \midrule
\rowcolor{lime} 37 & Transfusion: Predict the Next Token and Diffuse Images with One Multi-Modal Model & AI & \url{http://arxiv.org/abs/2408.11039} & 24/08/18-24/08/24 & 36 & 18.0 & - & - & - \\ \midrule
\rowcolor{lime} 38 & KAN: Kolmogorov-Arnold Networks & LG & \url{http://arxiv.org/abs/2404.19756} & 24/04/28-24/05/04 & 181 & 18.0 & - & - & - \\ \midrule
\rowcolor{Gray} 39 & Vision Mamba: Efficient Visual Representation Learning with Bidirectional State Space Model & CV & \url{http://arxiv.org/abs/2401.09417} & 24/01/14-24/01/20 & 385 & 17.9 & $\uparrow$N & - & - \\ \midrule
\rowcolor{Gray} 40 & Stable Video Diffusion: Scaling Latent Video Diffusion Models to Large Datasets & CV & \url{http://arxiv.org/abs/2311.15127} & 2023/11/19-2023/11/25 & 499 & 16,6
 & $\uparrow$N & - & - \\ \midrule

\bottomrule
\end{tabular}
    \caption{\textbf{Top 21-40} papers with their titles, primary categories (cs.XX), links, publication weeks, citation counts, z-scores, and trends in rankings compared to the top-40 lists in our previous report \protect\cite{zhang2023nllg},
    ranked by z-scores as of January 21, 2024. Row colors correspond to the publication time: \colorbox{white}{white} (2023/1/1-2023/6/30, added in \protect\arxivlargeppp{}), \colorbox{LGray}{light gray} (2023/7/1-2023/9/30, added in \protect\arxivlargepp{}), and \colorbox{Gray}{gray} (2023/10/1-2024/1/31, added in \protect\arxivlargep{}), and \colorbox{lime}{lime} (2024/1/31-2024/9/30, added in \protect\arxivlarge{}, i.e., this report).}
    \label{tab:top2140}
\end{table}

Of the recurring papers on our list, the highest position gains are achieved by Direct Preference Optimization (DPO) \cite{rafailov2024directpreferenceoptimizationlanguage}, a training method for human alignment (+7 ranks) and 3D Gaussian Splatting, a method enabling real time novel-view synthesis by infering volumetric representations of a scene (radiance fields) (+19 ranks)\cite{kerbl20233dgaussiansplattingrealtime}. Also, on rank 10, a survey about retrieval augmented generation (RAG) \cite{gao2024retrievalaugmentedgenerationlargelanguage} enters our list. It has already been published in December \se{2023,} 
which could indicate that more researchers are now working on this topic. Most ranks are lost by PaLM-E (-17) \cite{driess2023palmeembodiedmultimodallanguage} and Baichuan 2 (-14) \cite{yang2023baichuan2openlargescale}, which is likely caused by the introduction of newer, more competetive LLMs. Also, \textit{Visual Instruction Tuning} \cite{liu2023visualinstructiontuning} has lost 11 ranks, likely because the follow-up paper \textit{Improved Baselines with Visual Instruction Tuning} \cite{liu2024improvedbaselinesvisualinstruction} was introduced, which has now entered our list at rank 17. Further, the prompting technique Tree of Thoughts \cite{yao2023treethoughtsdeliberateproblem} has lost another 13 ranks, perhaps because of a costly usage and difficult setup in practical applications. It may also highlight a decrease of interest in prompting techniques.
Generally, among the primary categories of the top papers are distributed into 1 cs.GR, 1 cs.SE, 2 cs.AI, 8 cs.LG, 11 cs.CV and 17 cs.CL papers (see Figure \ref{fig:current_prev_top}). While there is no cs.CV paper amongst the top 10 anymore, their number increased by 2, which might show a slight focus shift towards computer vision. 
\se{Interestingly, while still dominating, cs.CL seems to be on the decline in favor of cs.CV and cs.LG; this indicates that more diversified foundation models across different AI subfields become a focus.
Figure \ref{fig:current_prev_top} also shows that there is a relatively strong increase in new papers in the top-20 and top-40 lists since our earlier reports; this may partly be explained by the relatively large time gap of 8 months since our last report. 
}

\begin{figure}[tb]
    \centering
    \begin{subfigure}[t]{0.49\textwidth}
    \includegraphics[width=\linewidth]{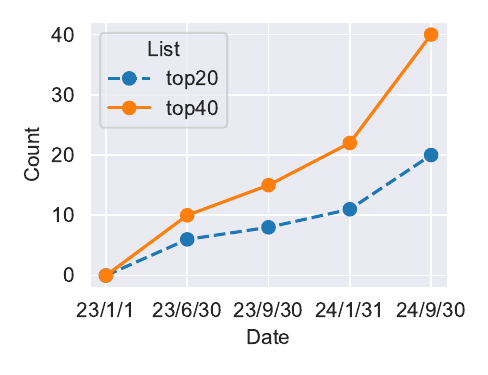}
    \caption{Inclusion of the current top 20 resp. top 40 papers in the previous reports' top 20 resp. top 40 papers.}
    \label{fig:top_inclusion}
    \end{subfigure}\hspace{.1cm}
    \begin{subfigure}[t]{0.49\textwidth}
    \includegraphics[width=\linewidth]{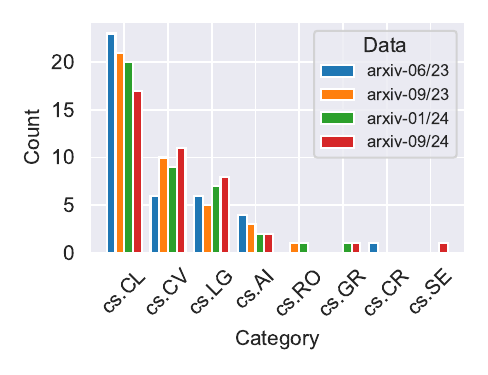}
    \caption{Distribution of arXiv categories among the current and previous reports' top 40 papers.}
    \label{fig:fine_cat_prev}
    \end{subfigure}
    \caption{Comparison of the current and previous top papers.}
    \label{fig:current_prev_top}
\end{figure}\FloatBarrier%
\section{\se{AI} \se{g}enerated content in scientific papers}
The advance of LLMs has made it simple to generate and paraphrase new texts. Hence, 
\se{researchers} 
also use LLMs in the process of writing scientific papers \cite{liang2024mappingincreasingusellms}. As this report's focus topic, we 
\se{extend} 
previous findings with a specific focus on \se{(i)} AI related sub-fields of arXiv and on 
\se{(ii)} 
top-40 papers. 

\subsection{Related work} Many works consider the detection of \se{AI} generated texts \cite{wu2024surveyllmgeneratedtextdetection}. Here, we give a brief overview of related work regarding the detection regarding scientific publications. 
\citet{liang2024mappingincreasingusellms} analyze papers from multiple online repositories, including arXiv and determine that the use of generated content indeed increased a few weeks after the release of ChatGPT. They use a corpus-level detection method \cite{liang2024monitoringaimodifiedcontentscale} that estimates the fraction $\alpha$ of \se{AI} generated text. To do so, they (1) construct fully human-written and AI-generated data sets, and (2) use maximum likelihood estimation to learn an estimator that predicts $\alpha$ for mixed distributions (from 1). For \textit{cs} papers on arXiv, they estimate that a fraction of over 17.5\% of all introduction sections and abstracts (combined into a single corpus) was generated in February 2024. Additionally\se{,} they find that the words \textit{realm,
intricate, showcasing, pivotal} are the most disproportionally used words by LLMs and displaying their frequency also highlights the increased use of generated content. 
\se{Contemporaneously with them},  \citet{geng2024chatgpttransformingacademicswriting} also propose a word frequency based method. They use arXiv abstracts 
published before the release of ChatGPT and rewrite them with the LLM to determine ground truth frequency changes. Then, they test different \se{m}odels to predict LLM impact and show their performance on unseen data.  
Interestingly, besides common LLM generated words, they also find that some words, like \textit{is} and \textit{are} are decreasing in frequency. \citet{gray2024chatgptcontaminationestimatingprevalence} 
specially focus\se{es} on the increasing use of specific words or word groups without learning an estimator. For example, the number of papers that use 
\se{two}
words 
\se{from} ``\textit{intricate, meticulous, meticulously, commendable}'' strongly increases. %
Another work specifically focusing on changes in word usage is \citet{kobak2024delvingchatgptusageacademic}. They explore the changes of word frequencies in PubMed abstracts since 2010 and show that the effects of the introduction of LLMs even outweights terms like \textit{pandemic} during Covid-19. As a result\se{,} they identify 319 style words with elevated usage since the release of ChatGPT.

A different line of work uses learned black-box LLM detection tools to quantify the usage of LLMs in academic writing. \citet{akram2024quantitativeanalysisaigeneratedtexts} uses an online AI detection tool to show increased use of LLM generated content in computer science papers on arXiv and \citet{DBLP:journals/apin/PicazoSanchezO24} explore different detectors for abstracts of multiple sources, but not in a 
\se{diachronic} 
comparison. 

By now\se{,} some subject specific works also explore specific domains based on word-frequencies, like dental research \cite{Uribe2024-ns} and astronomy \cite{astro}. Newer work shows that \se{AI} generated content  
is also used in scientific presentations and thereby also becomes a part of the human spoken language \cite{geng2024impactlargelanguagemodels}. This is also explored by \cite{yakura2024empiricalevidencelargelanguage} who analyze 20,000 Youtube channels and find increasing use of LLM terminology. 

Our work differs from these works by our specific focus on \se{four} selected 
\se{subfields}
of the arXiv computer science category. \cl{Notably, we assess whether the amount of AI generated content varies between the most influential papers and randomly chosen ones. This is an intriguing question as (1) on one hand AI usage may have improved the writing quality and have supported the high citation counts, but (2) on the other hand the authors of top papers may have the most resources to optimize human written texts and (3) not using AI may \se{(perhaps)} lead to a more distinctive writing style,  increasing the chance of the paper to be remembered and cited.}  
Also, while most of the works only consider the abstract and introduction, we consider a larger part of the papers. Additionally, we explore how the usage of ``LLM words'' from different sources impacts the evaluation.

\begin{figure}[tb]
    \centering
    \includegraphics[width=0.95\linewidth]{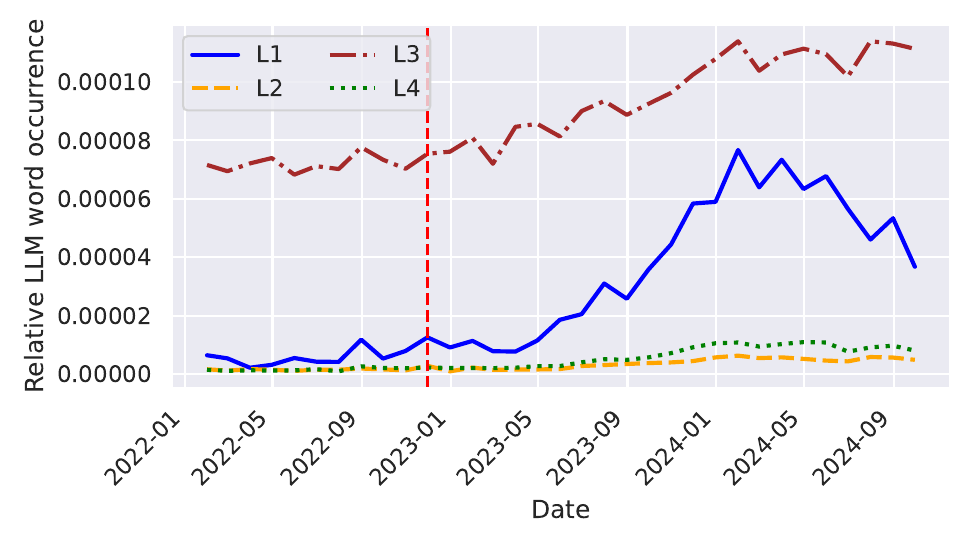}
    \caption{Relative word frequency per ``LLM-word'' list per month normalized by list length. List 1: \cite{liang2024mappingincreasingusellms}, List 2:  \url{https://github.com/FareedKhan-dev/Detect-AI-text-Easily}, List 3: \cite{kobak2024delvingchatgptusageacademic}, List 4: \cite{astro}.}
    \label{fig:list_diff}
\end{figure}

\subsection{Methodology}
Approaches of previous work can be grouped into 
\se{three}
categories \cite{kobak2024delvingchatgptusageacademic}: (1) detectors leveraging LLMs, (2) estimating mixture models and (3) zero-shot frequency based. Due to ease of setup, we apply two detectors leveraging LLMs: FastDetectGPT \cite{bao2024fastdetectgptefficientzeroshotdetection}, a method that compares the input text to sampled model outputs and Binoculars \cite{hans2024spottingllmsbinocularszeroshot}, a method that uses two closely related LLMs to determine whether the output of a model is more expected by a second model (cross-perplexity) than the human-written content. Additionally, we consider several word-lists for zero-shot frequency based detection: (1) the four top words by \cite{liang2024mappingincreasingusellms}, (2) a list compiled by user Fareed Khan on github\footnote{\url{https://github.com/FareedKhan-dev/Detect-AI-text-Easily}}, (3) the style words discovered by \cite{kobak2024delvingchatgptusageacademic} and (4) the 100 words identified by \cite{astro}. These are lists of words that people identified to be used more frequently by LLMs than by humans. 

\begin{figure}[tb]
    \centering
    \includegraphics[width=0.95\linewidth]{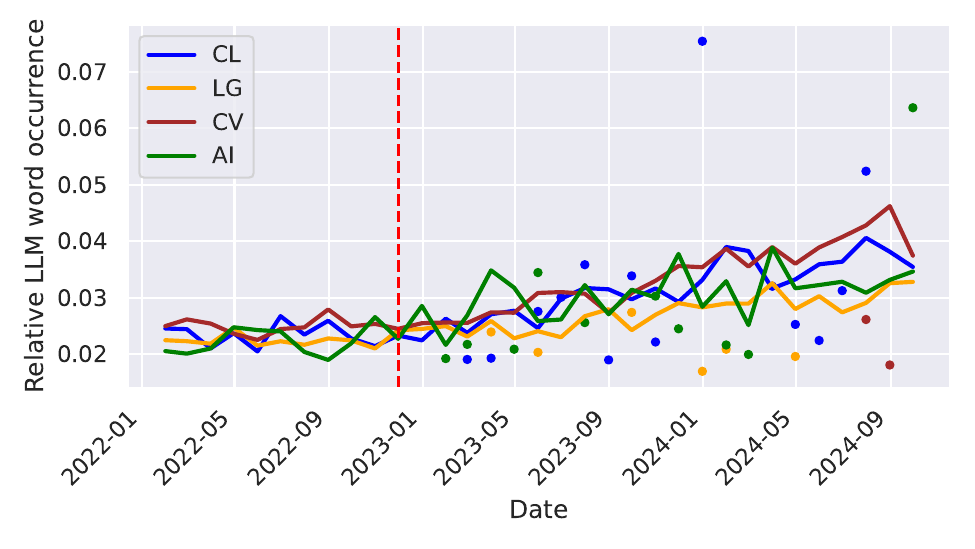}
    \caption{Relative word frequency per ``LLM-word'' per month and arXiv category. Frequencies per paper are aggregated by median. The points indicate the frequency of words for top-40 papers released in the respective months.}
    \label{fig:per_arxiv}
\end{figure}

\cl{\paragraph{Data}} To compare the top-40 papers to general papers on arXiv, we collect the PDF files of 10 random papers for each weekday from 2022/01/01 to 2024/09/30. Next, we use marker\footnote{\url{https://github.com/VikParuchuri/marker}} to extract the text content of these papers and the top-40 papers into markdown. This amounts to a total of 7144 papers (for 6 days only 9 papers where fetched). We use a larger time-span than the arXiv reports to be able to visualize changes since the release of ChatGPT (11/30/2022). The extracted data can be noisy and there is no clear way to extract only relevant content. As a simple heuristic, we use the first 40\% of each paper for the experiments. This should cut off the references and in many cases also most tables that are usually presented in the results sections. 

\begin{figure}[tb]
    \centering
    \includegraphics[width=0.95\linewidth]{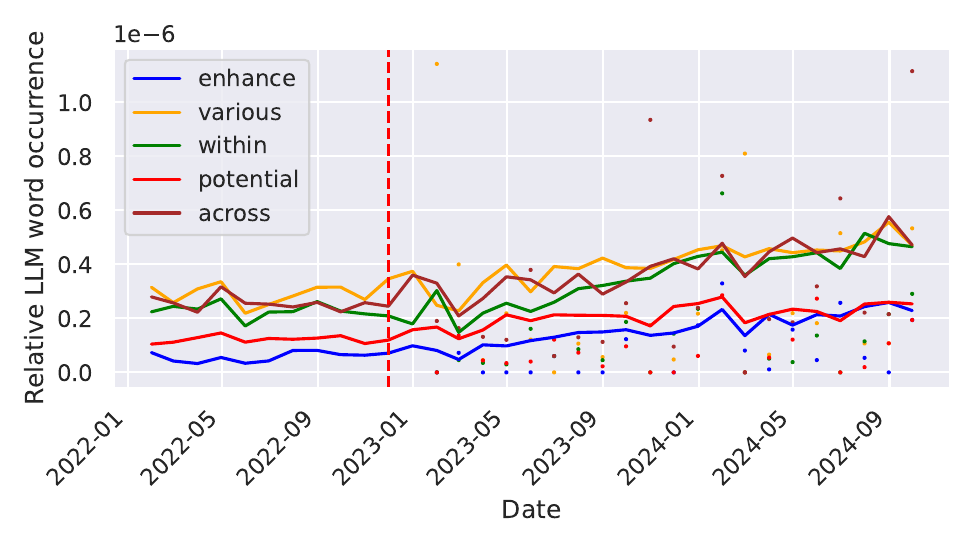}
    \caption{Top 5 words with the highest frequency change over time for random papers (line plot) and top-40 papers (scatter plot). For the selection, we average the relative frequencies of all papers in each month. Then, we take the median of the 11 months in 2022 till ChatGPTs release and of the 9 first months of 2024 and choose the words that show the highest increases.}
    \label{fig:top_words}
\end{figure}

\paragraph{Choosing the word lists}
To choose the most reasonable lists of ``LLM words'', for each list we determine the relative word frequency per month as follows:
\cl{
\begin{enumerate}
    \item Let \( C_{i,j} \) represent the count of the \( j \)-th ``LLM-word" in the \( i \)-th paper of the month. 
    \item Compute the total ``LLM-word" count for all papers and words of the month:
    \[C_{\text{total}} =  
    \sum_{i,j} C_{i,j}\]
    \item Determine the normalized relative word frequency:
    \[f_{\text{avg}} = \frac{C_{\text{total}}}{T \cdot N}\]
    where \( T \) is the total word count of all papers for the month and \( N \) is the number of ``LLM-words" in the list.
\end{enumerate}}

We plot the resulting frequencies over time in Figure \ref{fig:list_diff}. While all four word lists show a frequency increase over time, it is stronger for the words in List 1 and List 3. Notably, the word frequency in List 3 is already quite high before the release of ChatGPT, which shows that their words are more common in general, but still have increased since the release of ChatGPT. Another interesting finding is that List 1, which is composed of only 4 words is decreasing in frequency since early 2024. Perhaps this correlates with the release of newer \se{m}odels, prominently GPT4o in May 2024, that might use different training data. For the next experiments, we only use List 1 and List 3, due to their pronounced frequency change. Note, as a limitation, that we do not test for causality and some increases in word usage may be attributed to other factors like general language change and changes in the scientific fields due to the new technology. Causality may be given for some of the lists based on the methods their authors used.   

\subsection{Results}
\paragraph{Word-frequencies}
Next, for each paper in our dataset of 7144 papers, we determine the word-frequencies for List 1 and 3, sum them over all words and divide by the total word count of that paper. 
For each arXiv category, we select the median \cl{over all papers} per month as a robust measure. Then, we plot the resulting relative frequency of LLM words over time.
\cl{For each paper \( k \) in our dataset of  
$n=7144$
papers, we compute the relative frequency of ``LLM-words" as 
\begin{align*}
f_k = \frac{\sum_{j} C_{k,j}}{T_k}
\end{align*} 
where \( C_{k,j} \) is the count of the \( j \)-th word in List 1 or List 3 for paper \( k \), and \( T_k \) is the total word count of paper \( k \). For each arXiv category and month, we determine the median relative frequency as \( f_{\text{median}} = \text{median} \left( \{ f_k \} \right) \) for all \( k \) in the given category and month. Finally, we plot \( f_{\text{median}} \), the resulting median relative frequency of ``LLM-words" over time.}
For the top-40 papers that we determined in Section \ref{sec:top_papers},  
we do the same and add the results as scatterplots (see Figure~\ref{fig:per_arxiv}). \cl{That means, each point indicates the amount of AI generated content (based on word frequencies) in the current top-40 papers that were released in the respective month.} 

First, looking at the line-plots we can indeed see a steady increase of LLM word frequencies over time, confirming the previous findings that LLM-usage for academic writing is increasing. In our dataset, cs.CV has the highest frequencies for most months but this was also the case before the release of ChatGPT. cs.LG has the lowest frequencies. For the top-40 papers, most of them \cl{(22 out of 29 month+category combinations, i.e. the points in the graph)} are below the median frequencies of the random papers, 
which 
indicates that these papers \cl{may} use less \se{AI} generated content than others. 
However, especially the cs.CL category show\se{s} a wider variance, where some months lie  
\se{above}
the median frequency.

\begin{figure}[tb]
    \centering
    \includegraphics[width=0.95\linewidth]{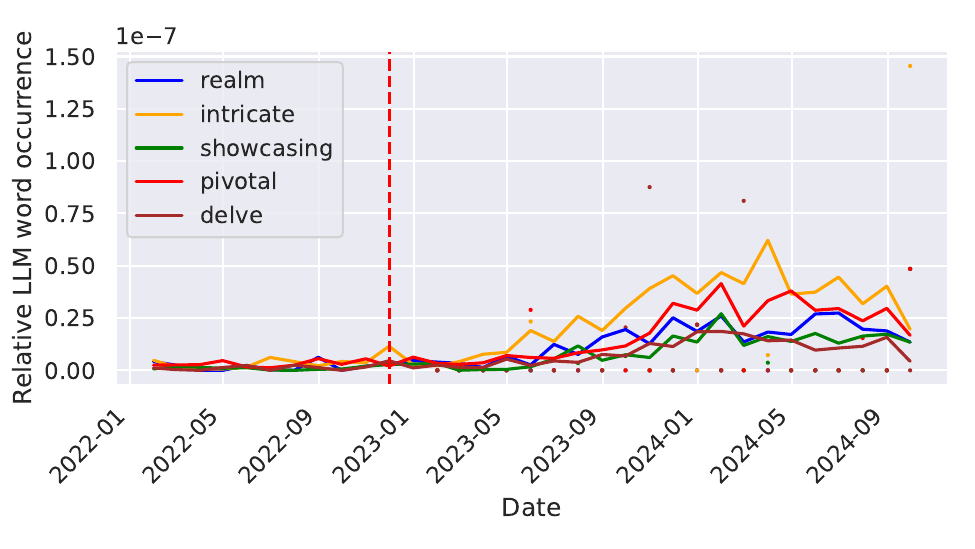}
    \caption{Previous top 4 words \cite{liang2024mappingincreasingusellms} \& ``delve'' for random papers (line plot) and top-40 papers (scatter plot).}
    \label{fig:top_words_previous}
\end{figure}

Further, Figure \ref{fig:top_words} shows the development of the top 5 words with the highest frequency change (of the selected word lists) from January 2022 to September 2024 for the random papers vs.\ the top\se{-}40 papers. All words are relatively common, so seeing their increase is surprising. Also, rather than a jump in usage, their frequencies are increasing steadily. Again, the top-40 papers 
rarely show 
higher frequencies than the random ones. \cl{For some months, the words ``within'' and ``across'' have high frequencies among the top-40 papers, as well.} 
In comparison previous key words used to determine LLM usage, like ``delve'' (see Figure \ref{fig:top_words_previous}) are declining in usage, i.e. contemporary LLMs seem to have been adapted.

\begin{figure}[tb]
    \centering
    \includegraphics[width=0.6\linewidth]{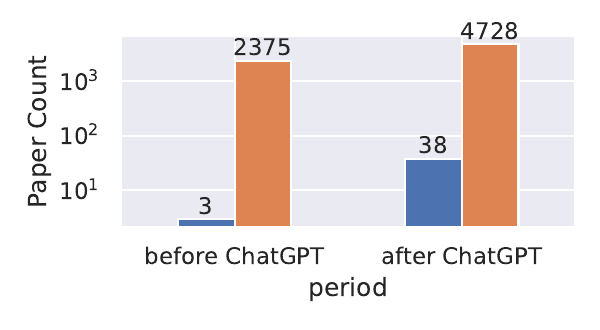}
    \caption{\se{AI g}enerated content detected by Binoculars before and after the release of ChatGPT.}
    \label{fig:bar}
\end{figure}

\paragraph{Detection Tools}
When we use FastDetectGPT and Binoculars, we can also identify very slight upward trends in the amount of generated content. For example\se{,} Figure~\ref{fig:bar} shows the absolute number of papers flagged as generated by Binoculars.\footnote{They propose two thresholds above which a text is generated. We use the lower one.} While there is a clear increase from 0.1\% to 0.8\% this type of detection seems to be too strict for our use-case as it assumes fully LLM-generated content. Plotting the detector scores as a line plot does not give conclusive results. Also (unsurprisingly), among the top-40 papers none was flagged as \se{fully AI} generated.

\FloatBarrier%
\section{Conclusion}\label{sec:conclusion}
We present the fourth NLLG arXiv report, highlighting the top 40 most influential papers in cs.CL, cs.AI, cs.LG and cs.CV. This includes 11 recently released highly cited papers and 7 further papers that gained influence over the past months. Notably, the focus lies on the creation of new foundation models and the search for newer, more efficient architectures. 
As this reports' special focus, we compare the prevalence of AI generated content in the top-40 papers with random papers selected from arXiv. We find that the top papers tend to contain less markers of AI use, indicating that they are written by humans to a larger degree. Additionally, we find that previous key-words to detect AI use are declining in frequency, calling into question the longevity of detection methods. To be more reliable, these tools would need to be updated on a model basis, e.g., by using fresh generated text for 
mixed-model based approaches. \FloatBarrier%

\section*{Acknowledgements}
The NLLG group gratefully acknowledges support from the Federal Ministry of Education and Research (BMBF) via the research grant ``Metrics4NLG'' and the German Research Foundation (DFG) via the Heisenberg Grant EG 375/5--1 
\se{and} 
the DFG Heisenberg grant EG 375/5--1.
\bibliographystyle{plainnat}
\bibliography{my} 

\end{document}